\documentclass[
aps,%
12pt,%
final,%
notitlepage,%
oneside,%
onecolumn,%
nobibnotes,%
nofootinbib,%
superscriptaddress,%
noshowpacs,%
centertags,
secnumarabic,]%
{revtex4}

\usepackage{amsfonts}
\usepackage{amssymb}
\usepackage{amsmath}
\usepackage{bm}
\usepackage{latexsym}

\newcommand{\beq}{\begin{eqnarray}}
\newcommand{\eeq}{\end{eqnarray}}
\newcommand{\be}{\begin{equation}}
\newcommand{\ee}{\end{equation}}
\def\la{\mathrel{\mathpalette\fun <}}
\def\ga{\mathrel{\mathpalette\fun >}}
\def\fun#1#2{\lower3.6pt\vbox{\baselineskip0pt\lineskip.9pt
\ialign{$\mathsurround=0pt#1\hfil ##\hfil$\crcr#2\crcr\sim\crcr}}}

\newcommand{{\SD}}{\rm SD}

\newcommand{\ver}{\mbox{\boldmath${\rm r}$}}

\newcommand{\vep}{\bm p}

\newcommand{\veR}{\mbox{\boldmath${\rm R}$}}

\newcommand{\lan}{\langle}
\newcommand{\ran}{\rangle}
\begin{document}
\title{The $X(6550), X(6900), X(7280)$ resonances as the  $nS, cc\bar c\bar c$ states}
\author{A.~M.~Badalian \\
NRC ``Kurchatov Institute" \\
Russia, Moscow}

\date{\today}
\begin{abstract}
Within the diquark-antidiquark model the masses of  the $0^{++}, cc\bar c\bar c$ resonances are calculated, using the expansion of the four-quark wave function
in the set of the hyperspherical functions. The interaction is defined via a universal pair-wise potential, which does not contain fitting parameters. The
resulting masses $M_4(nS)$ are shown to be very sensitive to the value of $c-$quark mass, chosen in relativistic string Hamiltonian, and
$m_c=1.24, 1.30, 1.43$ (in GeV) are considered. The choice of $m_c$, equal to the current mass, $m_c=1.245$ GeV,  yields three  $nS (n_r=0,1,2)$ states
in a very good agreement with the masses of the $X(6550), X(6900), X(7280)$ resonances, if the gluon-exchange interaction is totally neglected. This fact
indicates on a possible screening of the gluon-exchange interaction in the $cc\bar c\bar c$ system. For  $m_c=1.43$~GeV the ground state mass $M_4(1S)=6557$~MeV is obtained in agreement with experiment only if $\alpha_{\rm V}\cong 0.39(1)$ is used, however, in this case the masses of the $2S, 3S$ radial excitations exceed the masses of $X(6900), X(7280)$ by $\sim 100$~MeV.
\end{abstract}

\maketitle

\section{Introduction}

Recently the LHCb, CMS, ATLAS Collaborations have observed several resonances in the $J/\psi J\psi, J\psi \psi(2S)$ systems \cite{1,2,3}
(the summary of their results is also presented in \cite{4,5}). These experimental data are of a special importance for theory, because they
give an opportunity to test our understanding of the multiquark dynamics. The existence of the $QQ\bar Q\bar Q~(Q=c,b)$ resonances was
predicted long ago \cite{6,7,8}, where in \cite{8} the ground-state mass $M_4(1S)=6.50(1)$~GeV was obtained. Also the mechanism of
double $J/\psi$ production in proton-proton collisions was suggested  already in 2011 \cite{9}.  Therefore it is not surprising that
after observation of the $X(6550), X(6900), X(7280)$ resonances very large number of theoretical studies have been presented \cite{10,11,12,13},
including  comparison of results in different models. This comparison shows that predicted mass of the $0^{++}$ ground state varies in very wide range,
from the value $\sim 5.8$~GeV, i.e. below the $J/\psi J/\psi$ threshold \cite{9,14,15,16}, or close to the threshold \cite{11},
\cite{17}-\cite{20}, or $\sim 6.5$ GeV \cite{10,21,22,23}. In \cite{10} it was  also underlined that the gap between
$2^{++}$ and the ground state can change ten or more times, from 20 MeV up to 400 MeV, in different models. At the moment the reason of such
strong discrepancies is not clear and needs a special analysis.

Thus in the $cc\bar c\bar c$ system we face  the situation, which does not exist in charmonium, where within the same models
the charmonium spectrum is described in reasonable  agreement with experiment. In our paper we put the goal to understand how the
$cc\bar c\bar c$ spectrum depends on the most important parameters. In the case of charmonium the parameters of a static potential
$V_2(r)$ are well known due to studies of the static potential in the lattice QCD \cite{24,25} and the background perturbation theory \cite{26};
just these results were used in \cite{27} to define an universal potential.  In particular, the string tension of the confining potential
has the same value, $\sigma_2=0.180(2)$~GeV$^2$ at the distances $\la 1.2$~fm, as in leading Regge trajectory of light mesons. Also, on
the lattice it was established that the strength $F(r)=\frac{\partial V_2}{\partial r}$  satisfies the conditions: $r_0^2F(r_0)=1.65$ at
the point $r_0=0.455(6)$~fm and $r_1^2F(r_1)=1.0$ at the point $r_1=0.304(3)$~fm \cite{24,25,29}.

Also in the gluon-exchange (GE) potential the vector strong coupling $\alpha_{\rm V}(r)$ is known at small and large distances, both
in the momentum and the coordinate spaces \cite{28,29,30}, since the QCD constant $\Lambda_{\overline{MS}}$, as well as the vector
constant $\Lambda_{\rm V}$, is now determined  for $n_f=3,4,5$ \cite{28}. However, in the fully charm tetraquark the energy scale
$\mu$ of the strong coupling $\alpha_{\rm V}(\mu)$ is not a priori known, as well as its value at asymptotic. Correspondingly,
the value of the  $c-$quark mass, which is equal to the pole mass in charmonium, $m_c\cong 1.43(5)$~GeV, can differ  in
a four-quark system. In present studies of the $X_{4c}$ resonances  we take different values of  $m_c$ and the strong coupling, which factually
appear to be fitting parameters, to explain experimental data. For that we will use a relativistic version of the diquark-antidiquark model
\cite{8}  with the use of the relativistic string Hamiltonian (RSH) \cite{31,32} and keep expansion of the wave function in the set of the
hyperspherical functions \cite{33,34}. The  masses of the $0^{++}$ states with $n_r=0,1,2,3$ will be calculated with different  values of
$m_c=1.245, 1.30, 1.43$ (in GeV) and the strong coupling, and shown strong dependence of the tetraquark mass on these parameters.
The role and the scale of the strong coupling is  discussed.

\maketitle

\section{The charmonium $nF~(L=3)$ states}

Here we consider the $nF$ states in charmonium, which have some common features with the $0^{++}, cc\bar c\bar c$ states. Namely,
in the diquark-antidiquark model \cite{8}, where the wave function is expanded in the set of the hyperspherical functions and
the approximation of the minimal harmonic $K=0$  is used, the effective potential contains a ``centrifugal" term, similar to that in a
two-body system with the angular momentum $L=3$. This term appears even if all orbital momenta of a diquark, or antidiquark, and their relative orbital
momenta are equal zero; it is not produced by an orbital momentum, but is a  part of
the kinetic energy in the multidimensional system. Below we calculate the masses $M(nF,c\bar c)$ and compare them in the RSH and
the spinless Salpeter equation (SSE).

In the string picture \cite{27,31,32} the $c\bar c$ spectrum is defined by the RSH  $H_2(\rm str.)$,

\be
H_2(\rm str.) = \omega_2(nl) + \frac{m_c^2}{\omega_2(nl)} + \frac{\vep^2}{\omega_2(nl)} + V_2(r),
\label{eq.01}
\ee
where the potential $V_2 = \sigma_2 r + V_2^{(ge)}(r)$ has no fitting constants. It is the important point, because with
a large fitting constant, as in the Cornell potential \cite{35}, dynamical picture can be distorted.

To define the $c$-quark kinetic energy, the condition --  $\frac{\partial H_2(\rm str.)}{\partial \omega_2} = 0$ is imposed and it gives
\be
\omega_2(nl) = \sqrt{m_c^2 + \vep^2}.
\label{eq.02}
\ee
Then  $H_2(\rm str.)$ acquires the form of well-known spinless Salpeter equation (SSE):
\be
H_2(\rm SSE) = 2\sqrt{m_c^2 +\vep^2} + V_2(r); ~~ M_2(\rm SSE) = 2 \omega_2(nl) + \lan V_2\ran_{nl}.
\label{eq.03}
\ee
However, numerical solutions of the SSE are not always transparent, moreover, the  SSE wave function
needs a regularization at small $r$, thus introducing additional parameters. For that reason  in RSH  so-called
Einbein Approximation (EA) was suggested \cite{36,37}, where another condition,
\be
 \frac{\partial M_2}{\partial \omega_2} = 0,
\label{eq.04}
\ee
is imposed on the mass,
\be
M_2(nl) = \omega_2  + \frac{m_c^2}{\omega_2} + E_2(nl).
\label{eq.05}
\ee
In both approaches, the EA and SSE, the kinetic energies $\omega_2(nF)$ have close values, with a difference $\sim (10-20)$~MeV.
Also the $nF$ states have large sizes, $\ga 1.0$~fm  and in the $c\bar c$ interaction the linear potential (LP) $\sigma_2 r$ dominates.
For LP the condition (\ref{eq.04}) gives following equation,
\be
\omega_2^2(nF) = m_c^2 + \frac{1}{3} \sigma_2^{2/3} \zeta(nF)  \omega_2^{2/3}(nF).
\label{eq.06}
\ee
which has an analytic solution \cite{38}. For $\sigma=0.18$~GeV$^2$ and $m_c=1.43$~GeV the
following values of $\omega_2(nF): 1.674, 1.737, 1.796, 1.850$ (in GeV) $(n_r=0,1,2,3)$ are obtained. In (\ref{eq.06}) $\zeta(nF)$ are
the Airy numbers: $\zeta(1F)=5.051, \zeta(2F)=6.3322, \zeta(3F)=7.50465, \zeta(4F)= 8.5971$. The masses $M_2(\rm EA)$ (\ref{eq.05}) are
compared with the solutions of the SSE -  $M_2(\rm SSE)$ and in Table~\ref{tab.01} one can see small difference
between them, $\sim (10-30)$ MeV, for the $1F,2F,3F$ states .

The masses $M_2(nF)$ were also calculated in the linear+GE potential and even a smaller difference between $M_2(\rm EA)$ and $M_2(\rm SSE),
~ \la 10$~MeV ($n_r=0,1,2$) was obtained (see Table~\ref{tab.01}). Here in  the GE potential the two-loop coupling $\alpha_{\rm V}(r)$
with well established parameters was taken, namely, with the QCD vector constant $\Lambda_V=500$~MeV, which corresponds
to $\Lambda_{\overline{MS}}(n_f=3)= 339$ ~MeV, in full agreement with the
lattice analysis \cite{28}. Also in $\alpha_{\rm V}(r)$ the infrared regulator is taken from \cite{26,29}, which gives
$\alpha_{\rm V}(\rm asym.)=0.63$ at asymptotic.

\begin{table} [h!]
\caption{The spin-averaged masses $M_2(nF,c\bar c)$ in the einbein approximation (EA) and for the spinless Salpeter equation (SSE) (in MeV)
in the linear potential (LP) and linear plus GE potential ($m_c=1.43$~GeV, $\sigma_2=0.180$~GeV$^2$, $\alpha_{\rm V}(\rm asym.)=0.63$)}
\begin{center}
\label{tab.01}
\begin{tabular}{|c|c|c|c|c|}\hline
   potential &   LP       &  LP        & LP+GE    &  LP+GE  \\

  State    &   $M_2(\rm EA)$  & $M_2(\rm SSE)$ & $M_2(\rm EA)$  &$ M_2(\rm SSE)$ \\

  $1F$         &  4252     &  4245      &  4060   &   4052  \\
  $2F$        &  4594     &   4573  &  4425   & 4410  \\
  $3F$        &  4902     & 4870     &  4750   &  4744 \\
  $4F$        &  5188     & 5137     &  5047    &  5005 \\\hline
\end{tabular}
\end{center}
\end{table}

It is worth to note the following characteristic features  of the $nF$ states in charmonium: 1. Their sizes,  $R_2(nF)=\lan r\ran_{nF} = 0.92, 1.13, 1.33$
(in fm), are not very large; 2. In the SSE  the GE contribution to the mass is $\sim 190$~MeV  for the $M_2(1F)$ and 150 MeV for $M_2(3F)$, being
only by $\sim 10$ MeV larger than those in EA.

\maketitle

\section{The  $0^{++}, cc\bar c\bar c$ resonances. The general approach}

Here we develop the approach from \cite{8}, where in NR approximation the $0^{++}$ ground state mass, $M_4(1S) = 6.50(1)$~GeV,
was defined, using the expansion of the diquark-antidiquark wave function in the set of the hyperspherical functions \cite{33,34}.
However, in \cite{8} the static Cornell potential \cite{35} was used, which has large fitting constant, $C_0= - 0.84$~GeV, and extremely
large $m_c=1.84$ GeV was taken in NR Hamiltonian. Here we will use the RSH, which takes into account relativistic corrections,
and realistic $m_c$, equal to the current $c-$quark mass, $m_c=(1.27\pm 0.03)$~GeV, or the pole mass, $m_c\sim 1.43$~GeV.

In the four-quark system with the quark coordinates $\ver_i$ the RSH has the form, similar to that in charmonium (\ref{eq.01}),
\be
H_4(\rm str.) = \hat{T} + V_4(\ver_i),
\label{eq.07}
\ee
where the kinetic energy operator $\hat{T}$ is
\be
\hat{T} = \sum_i\left(\frac{\omega_i}{2} +\frac{m_c^2}{2\omega_i} + \frac{\vep_i^2}{\omega_i}\right).
\label{eq.08}
\ee
As in \cite{8}, this term can be rewritten via three Jacobi coordinates $\vec\xi_i$ and the coordinate
$\veR$ of the center of the mass:

\be
\vec\xi_1=\sqrt{\frac{m_Q}{2a}}(\ver_1 - \ver_2),~~ \vec\xi_2 = \sqrt{\frac{m_Q}{2a}}(\ver_1 + \ver_2  - \ver_3 - \ver_4), ~~\\
\vec\xi_3 =\sqrt{\frac{m_Q}{2a}} (\ver_3 - \ver_4).
\label{eq.09}
\ee
It gives
\be
\hat{T} =  \frac{m_Q}{8a}\frac{\partial^2 }{\partial \veR^2 } + \frac{m_Q}{2a} \Delta_{\xi}.
\label{eq.10}
\ee
Here in $\hat{T}$ and $\vec\xi_i$ there is an arbitrary scale parameter $a$, which does not change the excitation energies $E_4(nS)$ and
in our analysis we take $a=a_1=\frac{m_c}{2}$, which differs from the scale $a_2=m_c$, used in \cite{8} (see discussion below). This choice
$a_1=\frac{m_c}{2}$ seems to be preferable to perform  comparison with charmonium.

The Laplacian  $\Delta_\xi$ can be expressed via eight angular variables $\Omega_8$ (all definitions and details
are given in \cite{8}) and the hyper-radius $\rho,~\rho^2 = \vec\xi_1^2 +\vec\xi_2^2 + \vec\xi_3^2$,
\be
\Delta_\xi = \frac{1}{\rho^8} \frac{\partial }{\partial \rho} \left(\rho^8 \frac{\partial }{\partial \rho}\right) + \frac{\Delta_{\Omega}}{\rho^2} .
\label{eq.11}
\ee
The wave function $\Psi(r_{ij}, s_i, c_i)$ of the $0^{++}$ state can be presented as a sum of two terms:

$\Psi(r_{ij}, s_i, c_i) =\Phi_1(r_{ij}, s_i) \chi^{(\bar{3}3)}(c_i) + \Phi_2(r_{ij}, s_i) \chi^{(6\bar{6})}(c_i)$, where $r_{ij}, s_i, c_i$ are
the coordinate, spin and color variables. In the antisymmetric colour function $\chi^{(\bar{3}3)}$ the diquark (antidiquark) form a color
antitriplet (triplet) in the color space and in the symmetric function $\chi^{(6\bar{6})}$ a diquark (antidiquark) are in the sextet
(antisextet) state. The important point is that in  approximation of the minimal $K$ harmonic -- $K=0$  the wave functions $\Phi_i(\vec\xi_i) (i=1,2)$  are
defined by the equations with the same potential $V_{00}(\rho)$,

\be
V_{00}(\rho) = V_4^{(\bar{3}3)}(\rho)=V_4^{(6\bar{6})}=\frac{35}{32} \left( \sqrt{\frac{2a}{m_Q}}\sigma_2 \rho  - 4\sqrt{\frac{m_Q}{2a}}\frac{\kappa_2}{\rho}\right).
\label{eq.12}
\ee
Here the scale $a$ is not yet defined but from  the definition of the coordinates (\ref{eq.09}) one can see that the product
$\rho \sqrt{\frac{2a}{m_Q}}$ does not depend on the scale $a$. In the $\rho$-space the wave function,
\be
\psi(\vec\xi_i) = \frac{1}{\rho^3} \phi_0(\rho) u_0(\Omega), ~~ u_0(\Omega) = \sqrt{\frac{105}{32\pi^4}},
\label{eq.13}
\ee
satisfies the normalization condition,
\be
 8\int^\infty_0 {\rm d}\rho ~\rho^2~ |\phi_0(\rho)|^2  =1 .
\label{eq.14}
\ee
In the case of the $0^{++}$ state the eigenvalues of the diquark-antidiquark system, both in $\bar{3}3$ and $6\bar{6}$ representations, are defined by the equation,
\be
\frac{d\varphi_0}{d\rho^2} + \frac{2}{\rho}\frac{d\varphi_0}{d\rho} + 2a (E_4(nS) - W_{00}(\rho)) \varphi_0(\rho)  = 0 .
\label{eq.15}
\ee
with the effective potential $W_{00}(\rho)$,
\be
W_{00}(\rho) = V_{00}(\rho) + \frac{12}{2a\rho^2},
\label{eq.16}
\ee
which includes the ``centrifugal" term, as in the charmonium $nF (L=3)$ states. For the scale $a_1=\frac{m_c}{2}$~ the
potential $V_{00}(\rho)$ is
\be
V_{00}(\rho) = - \frac{\kappa_4}{\rho} + \sigma_4 \rho, ~~
\label{eq.17}
\ee
with
\be
\kappa_4 = \frac{35}{8} ~\kappa_2; ~~\sigma_4 = \frac{35}{32}~ \sigma_2,
\label{eq.18}
\ee
and for the standard $\sigma_2=0.18$~GeV$^2$ one has
\be
\sigma_4=1.09375~\sigma_2 = 0.1969~\rm GeV^2,
\label{eq.19}
\ee
and
\be
\kappa_4 = 4.375 ~\kappa_2.
\label{eq.20}
\ee

In section 2 we have discussed the parameters of  $\alpha_V(r)$, which are now known in the momentum and the coordinate spaces,
while in the $\rho-$space properties of $\kappa_4(\rho)$ were not yet studied on fundamental level. If one assumes that the behavior of
$\kappa_4(\rho)$ is similar to that of $\alpha_{\rm V}(r)$, e.g. changing $r$ by $\frac{\rho}{2}$, then the coupling has
the asymptotic freedom (AF) behavior at small $\rho$ and $\kappa_4\rightarrow \rm const.$ at large $\rho$. Then from the
equation (\ref{eq.20}), for typical $\alpha_V(\rm asym.)=0.60(4)$,  at asymptotic $\kappa_4(\rm asym.)=4.375~\kappa_2(\rm asym.)=
5.833~\alpha_V(\rm asym.)\cong 3.50(12)$, if there is no any physical  restrictions on the GE interaction in a four-quark system.
Here we assume that a screening of the $V_4^{(ge)}(\rho)$ potential is possible and impose a restriction on the size of the ground state:
$R_4(1S)\geq 0.80$~fm, i.e. the size cannot be too small, which leads to the restriction on the coupling,
$\kappa_4(\rm eff.)\leq 2.40$ (see  next sections).


\section{The $0^{++}, cc\bar c\bar c$ states in the linear potential}

As in charmonium, some nontrivial properties of the diquark-antidiquark system can be already seen in linear confining
potential, $V_4(\rm conf.)=\sigma_4 ~\rho$ (\ref{eq.19}). In NR approximation in Eq.~(\ref{eq.10}) the scale $a_1 = \frac{m_c}{2}$
for all states and from (\ref{eq.15}) the mass $M_4^{(nr)}(nS)$  is
\be
M_4^{(nr)}(nS) = 4 m_c + E_4(nS) = 4 m_c  + \left(\frac{\sigma_4^2}{m_c}\right)^{1/3} \zeta(nF),
\label{eq.21}
\ee
where $\sigma_4=0.1969$~GeV$^2$ corresponds to $\sigma_2=0.18$~GeV$^2$; the values of $M_4^{(nr)}(nF)$ are given in Table~\ref{tab.02}.

In the RSH every $nS$ state has to be considered separately with $a_1=\frac{\omega_4(nS)}{2}$ and in the LP the mass
$M_4(nS)$ is defined as

\be
M_4(nS) = 4\left(\frac{1}{2}\omega_4 + \frac{m_c^2}{2\omega_4}\right)  + \left(\frac{\sigma_4^2}{\omega_4}\right)^{1/3}  \zeta(nF),
\label{eq.22}
\ee
where $\omega_4(nS)$ is determined by the condition $\frac{\partial M_4}{\partial \omega_4}=0$
(einbein approximation), which gives the equation,

\be
\omega_4^2 = m_c^2 + \frac{1}{6}~ \sigma^{2/3}~ \zeta(nF)~\omega_4^{2/3}.
\label{eq.23}
\ee
With $\sigma_4=0.1969$ GeV$^2$ (\ref{eq.19}) and $m_c=1.43$~GeV their values  are
\be
\omega_4(1S) = 1.557,~ \omega_4(2S) = 1.591, ~\omega_4(3F)= 1.621, ~\omega_4(4F) =1.650~ (\rm in~ GeV).
\label{eq.24}
\ee
Note that the solutions of the equation (\ref{eq.23}) can be written in analytic form \cite{38}.
The values of $\omega_4(nS)$ in $cc\bar c\bar c$ system appear to by  $\sim 150$~MeV smaller than $\omega_2(nF)$
in charmonium, given in section 2. Knowing $\omega_4(nS)$, the masses $M_4(nS)$ were calculated according to ({\ref{eq.22})
for $m_c=1.43$~GeV and 1.30 GeV. Their numbers are given in Table~\ref{tab.02}, where one can see strong difference
between them, $\sim 400$~MeV. However, in the EA, due to relatively small values of $\omega_4(nS)$, the masses $M_4(nS)$ occur to be
only by (30-60) MeV smaller than in NR case.

\begin{table}
\caption{The masses of the $0^{++}, cc\bar c\bar c$ states,  $M_4^{(nr)}(nS)$ and $M_4(nS)$ (in GeV) in the LP with $m_c=1.43$~GeV and 1.30 GeV, $\sigma_4=0.1969$~GeV$^2$}
\begin{center}
\label{tab.02}
\begin{tabular}{|c|c|c|c|c|}\hline

   $M_4$         & $M_4^{(nr)}(nS)$ & $M_4(nS)$ &     $M_4^{(nr)}(nS)$  &   $M_4(nS)$ \\
   $m_c$   &  1.43       &    1.43        & 1.30    &  1.30      \\

  state           &              &               &        &   \\
  $1S$         & 7.235          & 7.215      &    6.766  &    6.738  \\

  $2S$          & 7.620         & 7.588       & 7.164   &7.120  \\

  $3S$          &  7.971       &  7.927      & 7.527  & 7.468 \\

  $4S$         &  8.299      & 8.241      &  7.866   &7.789  \\\hline
\end{tabular}
\end{center}
\end{table}
Thus in LP in both cases, with $m_c=1.43$~GeV and 1.30 GeV,  the ground state mass $M_4(1S)$ is significantly larger than
that of the $X(6550)$ resonance, which is considered as the ground $0^{++}$ state.

Surprising result was obtained with $m_c$ equal to the current mass, $m_c=1.245$~GeV. In this case $M_4(1S), M_4(2S), M_4(3S)$
occur to be in very good agreement with experimental masses of the $X(6550), X(6900), X(7280)$ (see Table~\ref{tab.03})
and therefore $X(6900), X(7280)$ can be interpreted as first and second radial $0^{++}$ excitations. Calculated masses, $M_4(1S,\rm th.)=6542$~MeV,
$M_4(2S,\rm th.)=6930$~MeV, and $M_4(3S)=7.282$~MeV coincide with the CMS data within small experimental error \cite{2,5}. Such a
good agreement seems not to be occasional and can be interpreted as possible suppression of the GE interaction in the $cc\bar c\bar c$ system.
Note that the choice of $m_c$, equal to the current mass, was also used in the QCD sum rules approach in \cite{39}, where
the mass of the $X(6550)$ resonance, equal to $(6570\pm 55)$~MeV was obtained.
\begin{table}
\caption{The masses $M_4(nS,\rm th.)$ (in MeV) of the $0^{++}, cc\bar c\bar c$ states in the linear potential ($\sigma_4=0.1969$~GeV$^2, m_c=1.245$~GeV)
and the masses of the $X(6500), X(6900), X(7280)$ resonances from \cite{1}--\cite{5}}
\begin{center}
\label{tab.03}
\begin{tabular}{|c|c|c|c|c|}\hline

state  & $M_4(nS,\rm th.)$ & the ATLAS data \cite{3,4}  & the CMS data \cite{2,5} & the LHCb data \cite{1}\\

  $1S$   &   6542        &$ 6620 \pm 30^{+20}_{-10}    $ &  $ 6552 \pm 10\pm 12$ &     \\

  $2S$    &  6930        &$ 6870 \pm 30^{+60}_{-10}    $ & $ 6927 \pm 9 \pm 5    $ & $ 6905 \pm 11 \pm 7$ \\

  $3S$    &  7282         & $ 7220\pm 30^{+20}_{-30}    $  & $7287 \pm 19 \pm 5     $&       \\

  $4S$   &   7.608       &           &    &       \\\hline
\end{tabular}
\end{center}
\end{table}

In conclusion we underline several characteristic features of the $0^{++}, cc\bar c\bar c$ states in LP.
\begin{description}
\item{1.} The sizes $R_4(nS)= \lan\rho \ran_{nS}$  practically coincide with $R_2(nF,c\bar c)$ of the
$nF$ charmonium states: $R_4(1S)=1.0,~R_4(2S)=1.22,~ R_4(3F)=1.43$~(in fm) ($m_c=1.43$~GeV).

\item{2.} The $c$-quark kinetic energies $\omega_4(nS)$ are by (150-250) MeV smaller than $\omega_2(nF)$ in charmonium.
Their values: $\omega_4(1S)=1.557,~ \omega_4(2S)= 1.591,~ \omega_4(3S)=1.621$ ~(in GeV) ($m_c=1.43$~GeV,~ $\sigma_4=0.1969$~GeV$^2$)
can be compared with $\omega_2(1S)=1.674,~\omega_2(2S)=1.737,~ \omega_2(3S)=1.795$ (in GeV) with
$\sigma_2=0.18$~GeV$^2$ in charmonium.

\item{3.} In the considered approach  the mass $M_4(4S)\cong 7600$~MeV is predicted.

\item{4.} The masses  $M_4(nS)$ strongly depend on the chosen value of $m_c$, but to make final conclusion on a preferable $m_c$
the GE potential has to be taken into account.
\end{description}
\section{The $0^{++}$ states in linear + gluon-exchange potential}

In a four-quark system there are some difficulties in definition of the GE coupling $\kappa_4$. For example, in diquark-antidikuark
model \cite{8}, where the wave function is expanded in a set of the hyperspherical functions  \cite{33,34} and approximation of
the minimal harmonic is used, the coupling $\kappa_4$ is proportional to conventional
$\kappa_2=\frac{4}{3}\alpha_V$ with large coefficient 4.375 (\ref{eq.20}). If there is no any restrictions on the
value of $\kappa_4$, then at large $r$ $\kappa_4(\rm asym.)$ reaches large value $\cong 3.50$, which corresponds to typical
$\kappa_2(\rm asym.)\cong 0.80$ in charmonium. Note that restriction on $\alpha_{\rm V}(\rm asym.)\cong 0.60(4)$ was proved in
phenomenological analysis of meson spectra (light, heavy, and heavy-light) and in the background perturbation theory \cite{26},
where at large $r$ $\alpha_V$ is frozen due to presence of the infrared regulator, $M_B^2\cong 2\pi\sigma$
($M_B=1.06\pm 0.11)$~GeV \cite{29}). One cannot exclude that a restriction on the $\kappa_4$ also exists but this fundamental problem
is not yet solved. For that reason here we will use different values of $\kappa_4$.

We assume that at small $\rho$ the coupling $\kappa_4(\rho)$ depends on $\rho$ as in  $\alpha_V(r)$ and use
a correspondence $\rho^2\sim 4r^2$. Then the AF behavior of $\kappa_4(\rho)$ takes place only up to distances
$\la 0.3 $~fm and at larger $r$ $\kappa_4$ quickly approaches a constant, called an effective coupling. We will impose a restriction
on $\kappa_4(\rm eff.)$, using as a criterium the size of the ground state, namely, the size $R_4(1S)$ is supposed to be $\ga 0.80$~fm,
or larger than doubled size of $J/\psi$.

We use here, as before, the scale $a_1=\frac{\omega_4(nS)}{2}$, specific for a given $nS$ state, and define the kinetic energy  from
the condition $\frac{\partial M_4}{\partial \omega_4} = 0$, imposed on the mass,

\be
M_4(nS) = 2\left(\omega_4 +\frac{m_c^2}{\omega_4}\right) + E_4(nS),
\label{eq.25}
\ee
where the parameter $\omega_4$ has to be singled out in all terms of $E_4(nS)$,

\be
E_4(nS)=\sigma_4^{2/3}~\zeta(nF)~\omega_4^{-1/3} -  \kappa_4 ~\sigma_4^{1/3}~ y^{-1}~\omega_4^{1/3}.
\label{26}
\ee
Here the dimensionless matrix element ( the number) $y^{-1}(nS)$ is defined as  $\lan\rho^{-1}\ran_{nS} = \sigma_4^{1/3}~\omega_4^{1/3}~ y^{-1}(nS)$.
The values of $y^{-1}(nS)$ depend on $\kappa_4$ and for $\kappa_4=0.84$  the numbers $y^{-1}(nS)$ are following: 0.341(1),
0.296, 0.266, 0.242 ~$(n_r=0,1,2,3)$; they weakly change under small variations of $\kappa_4$ in the range $0.84\pm 0.14$. Then

\be
\omega_4^2(nS) =  m_c^2  + \frac{\sigma_4^{2/3}}{6}~\zeta(nF)~\omega_4^{2/3}  +
  \frac{\kappa_4~ \sigma_4^{1/3}~ y^{-1}(nS)}{6}~ \omega_4^{4/3}.
\label{eq.27}
\ee

Now we consider two cases: with relatively small $\kappa_4(\rm eff.)=0.84$ (or $\alpha_V=0.144$) and  $\kappa_4(\rm eff.)=2.275$
(or $\alpha_V=0.39$). In both cases $\sigma_4=0.1969$~GeV$^2$ and $m_c=1.43$~GeV or 1.30~GeV.

The case A. $\kappa_4(\rm eff.)=0.84$.

The relation  $\kappa_4(\rm eff.)=4.375~ \kappa_2(\rm asym.)=0.84$ corresponds $\alpha_V(\rm eff.)=0.144$,
or $\kappa_2(\rm eff.)=0.192$.  This number $\kappa_4=0.84$ just coincides with typical $\kappa_2(\rm asym.)\cong 0.80(4)$
in charmonium, but is significantly smaller than $\kappa_4(\rm max.)\cong 3.50$. Such decreasing of the coupling $\kappa_4(\rm eff.)$
implies a strong screening of the pair-wise GE potentials inside the four-quark system, e.g. inside a compound bag. There is
also another reason why $\kappa_4(\rm eff.)$ could be small - a true mass scale $\mu$ of $\kappa_4(\mu)$
is not established in $cc\bar c\bar c$ system: it may be $m_c$, or $2m_c$, or even larger.

With $\kappa_4(\rm eff.)=0.84,~\sigma_4=0.1969$~GeV$^2$ and $m_c=1.30$~GeV calculated $\omega_4(nS)$ are following:
1.449, 1.482, 1.511, 1.541 (in GeV). Then from the equations (\ref{eq.25}) and (\ref{eq.15}) one obtains the masses of the $0^{++}$ states:
$M_4(1S)=6559, M_4(2S)=6957, M_4(3S)=7.323$~(in MeV) (see Table~ \ref{tab.04}). One can see that for the ground state $M_4(1S)$ agrees with
the experimental mass of the $X(6550)$ resonance,  $M(X(6550))=6552(22)$~MeV.  Also the masses of the $2S,3S$ radial excitations are
in reasonable agreement with those of the $X(6900), X(7280)$ resonances, being only by $\sim 40$~MeV larger. For all
three states not small GE contributions, $\delta^{(ge)}(nS) =- 187, -0.163,-145$~(in MeV) are obtained.
Thus one can conclude that with $m_c=1.30$~GeV and small $\kappa_4(\rm eff.)=0.84~(\alpha_V(\rm eff.)=0.144)$, one obtains
the masses of the $0^{++}, cc\bar c\bar c$ states in reasonable agreement with experiment.

Situation is different if a larger $m_c=1.43$~GeV is taken. In this case from (\ref{eq.27}) the kinetic energies are following,
$\omega_4(nS)= 1.575, 1.606,~1.635, 1.662$ (in GeV) and then from (\ref{eq.15}) one finds the excitation energies $E_4(nS)$,  which allow to
define the masses $M_4(nS)$ from (\ref{eq.25}), given in Table~\ref{tab.04}. One can see that with $m_c=1.43$~GeV the ground state mass,
$M_4(1S)=6995$ MeV appears to be by $\sim 450$~MeV larger than that of the $X(6550)$ resonance, although
it is by 220 MeV smaller than $M_4(1S)$  in LP (see Table~\ref{tab.02}). Also too large masses $M(2S)$ and $M(3S)$ are obtained
for radial excitations (see Table~\ref{tab.04}). This result means that in the case with $m_c=1.43$~GeV much stronger GE interaction is needed.

\begin{table}
\caption{The masses $M_4(nS)$ (in MeV) of the $0^{++}, cc\bar c\bar c$ states in the LP+GE potential with $m_c=1.30$ GeV and
          1.43 GeV, $\kappa_4(\rm eff.)=0.84,~ \sigma_4=0.1969$~GeV$^2$}
\begin{center}
\label{tab.04}
\begin{tabular}{|c|c|c|c|}\hline
  $m_c$ ~(in GeV)    &  1.43    & 1.30  & the CMS data \cite{2,5}\\

  $M_4(1S)$            &   6995       & 6559 &  6555(22)\\
  $M_4(2S)$            &   7381       & 6957   & 6927(14) \\
 $M_4(3S)$              & 7723       & 7323  & 7287 (24)    \\\hline
\end{tabular}
\end{center}
\end{table}

Our analysis shows that in the linear + GE potential with $\kappa_4=0.84$ the quark kinetic energies increase by only $(10-15)$~MeV as compared
to their values in the LP, while in charmonium with the same $\kappa_2=0.84$  $\omega_2(nS)$ increase  by $\sim (30-50)$~MeV, i.e.
relativistic effects are larger in charmonium.

\section{The masses $M(nS)$ in the gluon-exchange potential with $\kappa_4(\rm eff.)=2.275$}

In preceding section with the parameters $m_c=1.43$~GeV and $\kappa_4=0.84$ we have obtained the ground state mass $M_4(1S)=6995$ MeV,
which is by 430 MeV larger than experimental number. Here we take a larger $\kappa_4(\rm eff.)=2.275$ and show that in this case
$M_4(1S)$ agrees with data (in some sense this value of $\kappa_4$ can be considered as a fitting parameter). As the first step we use (\ref{eq.27})
to define the quark  kinetic energies $\omega_4(nS)$ and the following values: 1.611, 1.638, 1.663,~1.689~$(n_r=0,1,2,3)$ (in GeV)
were calculated. In (\ref{eq.27}) we have introduced again  the dimensionless numbers $y^{-1}(nS)$, defined via $\rho^{-1}$ as
$(\omega_4(nS) \sigma_4(nS))^{-1/3}\lan\rho^{-1}\ran_{nS} = y^{-1}_{nS} =  0.40(1), 0.33(1), 0.285(5), 0.269(5) ~(n_r=0,1,2,3)$.
These numbers depend on the value of $\kappa_4$ and can be considered as constants for the coupling in the range $\kappa_4\cong (2.2\pm 0.2)$.
Also these $y^{-1}(nS)$ occur to be only by $\sim 10\%$ larger than those for $\kappa_4=0.84$, presented in last section.

With  $\omega_4(1S)=1.611$~GeV and the effective potential,
\be
W_{00}(\rho) = \sigma_4 ~\rho - \frac{\kappa_4}{\rho} + \frac{12}{\omega_4~\rho^2},
\label{28}
\ee
in (\ref{eq.15}), one defines $E_4(1S)$ ($\sigma_4=0.1969$~GeV$^2, \kappa_4=2.275,~m_c=1.43$~GeV). Then from (\ref{eq.25}) the ground state mass
$M_4(1S)=6557$~MeV is obtained, which is in good agreement with the mass of the $X(6550)$ resonance
(see Table~\ref{tab.05}). However, in this case the masses $M(2S), M(3S)$ of the radial excitations appear to be by
(80-100)~MeV larger than experimental values.

In Table~\ref{tab.05} we give the sizes of the $nS$ states,  $R_4(nS)= \lan\rho\ran_{nS}$, which are not large, moreover,
$R_4(1S)=4.2$~GeV$^{-1}=0.83$~fm coincides with the doubled r.m.s. of $J/\psi$. This size $R_4(1S)\geq 0.80$~fm could be used
as a criterium to restrict the value of $\kappa_4(\rm eff.)$, i.e. if $m_c=1.43$~GeV, the coupling has to be frozen,
with the number $\kappa_4(\rm eff.)\cong 2.3$, otherwise the size of the ground state is becoming too small.

\begin{table}
\caption{ The sizes $R_4(nS)$ (in fm) and the masses $M_4(nS)$ (in MeV) in the LP+GE potential with $\kappa_4(\rm eff.)=2.275, m_c=1.43~ GeV$}
\begin{center}
\label{tab.05}
\begin{tabular}{|c|c|c|c|} \hline

          &$R_4(nS)$   & $M_4(nS)$  & the CMS data  \cite{2,5}\\

  $1S$    & 0.82  & 6550(10)      & 6555(22) \\
  $2S$    & 1.05 &  7023         & 6927(14)  \\
  $3S$    &  1.29   & 7362      & 7287(24)  \\
  $4S$    & 1.50   & 7702      &    \\\hline

\end{tabular}
\end{center}
\end{table}

\section{Conclusions}

In our paper we have studied the masses and the sizes of the $0^{++}, cc\bar c\bar c$ states with the use of the RSH in
the einbein approximation, where the structure of the mass formula is simple and very clear.
Our analysis was done using different values of the $c-$quark mass and the strong coupling $\kappa_4$ (or $\alpha_{\rm V}$), and
shown that there exists strong correlation between  $m_c$ and $\alpha_{\rm V}$ chosen to reach agreement with experimental data.
Surprisingly, the best agreement with experimental data on three resonances,  $X(6550), X(6900), X(7280)$,  was obtained in the
linear confining potential (the GE potential was totally excluded) with  $m_c=1.245$~GeV, equal
to the current mass. In this case the ground state mass $M(1S)=6.542$~GeV, as well as the masses of the radial excitations,
$M(2S)=6930$~MeV and $M(3S)=7282$~MeV, agree with data within experimental error; for the $4S$ state  $M_4(4S)=7600$~MeV is predicted.
This result can be interpreted as possible strong screening (suppression) of the GE interaction in a four-quark system. For that reason
the value of the strong coupling in spin-spin splitting is not also clear and here we calculated  only spin-averaged masses. Also

\begin{enumerate}

\item If $m_c=1.30$~GeV and the GE potential with small $\kappa_4=0.84$ is taken, then
 $M_4(1S)=6.559$~MeV is in agreement with mass of the $X(6550)$ resonance and the masses $M(2S), M(3S)$ of the radial excitations
 occur to be only by (30-50) MeV larger than those of the $X(6900), X(7280)$ resonances.

\item  The case with $m_c=1.43$~GeV, which provides  precision description of the charmonium spectrum, is studied in details.
For the ground state mass $M(1S)$ one can reach agreement with experiment, only if the coupling $\kappa_4\cong 2.30$ (or $\alpha_V\cong 0.39$)
is used. However, in this case $M(2S), M(3S)$ appear to be by $\sim 100$~MeV larger than experimental masses.

\item With large $\kappa_4=2.275$ and $m_c=1.43$~GeV a small size of the ground state, $R_4(1S)=0.83$~fm, was obtained. This size
is equal to doubled r.m.s. of $J/\psi$ and for that reason  it seems reasonable to restrict the value of the strong coupling
imposing the condition: $\kappa_4\la 2.3$, in order not to have too small $R_4(1S)$.

\item Important question what is the true mass scale of the strong coupling $\kappa_4$ in the $cc\bar c\bar c$ system, needs a special analysis
on fundamental level.

\end{enumerate}

I am very grateful to Prof. Yu. A. Simonov for useful discussions.

\end{document}